\documentclass[sigconf, authorversion=true]{acmart}

\setcopyright{none}

\usepackage{outlines}

\usepackage{subcaption}

\usepackage[]{hyperref}

\usepackage{pgfplots}
\pgfplotsset{compat=1.18}
\usepackage{filecontents}

\AtBeginDocument{%
  \providecommand\BibTeX{{%
    \normalfont B\kern-0.5em{\scshape i\kern-0.25em b}\kern-0.8em\TeX}}}

\copyrightyear{2025}
\acmYear{2025}
\setcopyright{acmlicensed}\acmConference[UCC '25]{2025 IEEE/ACM 18th International Conference on Utility and Cloud Computing}{December 1--4, 2025}{Nantes, France}
\acmBooktitle{2025 IEEE/ACM 18th International Conference on Utility and Cloud Computing (UCC '25), December 1--4, 2025, Nantes, France}
\acmDOI{10.1145/3773274.3774269}
\acmISBN{979-8-4007-2285-1/2025/12}

\usepackage{todonotes}

\begin{document}

\title{
  Modeling Anomaly Detection in Cloud Services: Analysis of the Properties that Impact Latency and Resource Consumption
}

 \author{Gabriel Job Antunes Grabher}
 \orcid{0000-0001-9415-7591}
 \affiliation{%
   \institution{Université Grenoble-Alpes, CNRS, Inria, Grenoble INP, LIG}
   \city{Grenoble}
   \country{France}
 }
 \email{gabriel.job-antunes-grabher@univ-grenoble-alpes.fr}

 \author{Fumio Machida}
 \orcid{0000-0001-9779-983X}
 \affiliation{%
   \institution{University of Tsukuba}
   \city{Tsukuba}
   \country{Japan}
 }
 \email{machida@cs.tsukuba.ac.jp}

 \author{Thomas Ropars}
 \orcid{0000-0002-9461-1165}
 \affiliation{%
   \institution{Université Grenoble-Alpes, CNRS, Inria, Grenoble INP, LIG}
   \city{Grenoble}
   \country{France}
 }
 \email{thomas.ropars@univ-grenoble-alpes.fr}


\renewcommand{\shortauthors}{G.J.A. Grabher, F. Machida, and T. Ropars}

\begin{abstract}


  Detecting and resolving performance anomalies in Cloud services is crucial for maintaining desired performance objectives. Scaling actions triggered by an anomaly detector help achieve target latency at the cost of extra resource consumption. However, performance anomaly detectors make mistakes. This paper studies which characteristics of performance anomaly detection are important to optimize the trade-off between performance and cost. Using Stochastic Reward Nets, we model a Cloud service monitored by a performance anomaly detector. Using our model, we study the impact of detector characteristics, namely precision, recall and inspection frequency, on the average latency and resource consumption of the monitored service. Our results show that achieving a high precision and a high recall is not always necessary. If detection can be run frequently, a high precision is enough to obtain a good performance-to-cost trade-off, but if the detector is run infrequently, recall becomes the most important.

 \end{abstract}




\maketitle

\section{Introduction \label{sec:intro}}

Cloud computing plays a major role in deploying and maintaining modern web-services applications~\cite{seemakhupt2023cloud}. Cloud infrastructures provide the necessary resources and flexibility to implement high availability and dynamic scaling techniques required by such services. However, correctly deploying and configuring Cloud services is challenging. Incorrect resource allocation and contention on shared resources are important sources of performance anomalies~\cite{lee2024tale, wydrowski2024load, iorgulescu2018perfiso}, which can negatively impact latency or lead to service crashes. Such problems are further amplified by the significant workload variations that a service can experience~\cite{shi2023nodens, cho2020overload}. The main technique used in Cloud environments to deal with such issues is scaling out (also called horizontal scaling), which implies creating more replicas of a service to improve performance~\cite{rzadca2020autopilot}. Hence, operating such services requires optimizing a trade-off between performance and resource consumption.



Cloud services are complex and performance anomalies, that is, deviations from the \emph{expected} performance under the current operating conditions, can be difficult to detect~\cite{rzadca2020autopilot}. Several techniques have been proposed to detect and correct performance anomalies~\cite{gan2021sage, qiu2020firm, chow2022deeprest, li2021multivariate, ghosh2024fast}. However, detectors can make mistakes that lead to inappropriate resource allocation. Existing solutions to detect anomalies differ in terms of achieved recall (proportion of actual anomalies that the detector correctly identifies), precision (proportion of detected anomalies that are actual anomalies), and inspection frequency (how frequently the detector can inspect the state of the service), among other things. These properties are important because they determine how fast and how accurately a system can react to an anomaly. However, it is difficult to determine which properties to tune to obtain the best trade-off between performance and resource consumption for the monitored service.


One solution to compare trade-offs offered by different solutions is to run experiments. However, Cloud software stacks and Cloud applications are very complex~\cite{gan2019open,luo2021characterizing,wydrowski2024load}. Collecting large amounts of monitoring data~\cite{lee2024tale} and training complex detectors is time consuming, especially for detectors based on deep learning. As an alternative, one can build a model of the system~\cite{ghosh2010end} and study through this model how different parameters influence its behavior.

This paper proposes a model to study a Cloud service monitored by a performance anomaly detection system, where performance issues are handled by scaling out. Based on this model, we conduct an analysis of the impact of the main characteristics of anomaly detectors (recall, precision, inspection frequency) on the trade-off between performance, measured as the latency of the service, and resource consumption. We focus on latency as a performance metric because latency is often considered as an important criteria for specifying Service-Level Objectives (SLOs). The proposed model is based on Stochastic Reward Nets (SRNs)~\cite{trivedi2001probability}. Since they allow for assigning rewards to model states, SRNs enable us to estimate properties such as latency or resource consumption.



Using the proposed model, we study the impact of the characteristics of anomaly detectors on the latency-cost trade-off that can be obtained when operating a Cloud service. Instantiating the model with realistic values from Cloud systems, we compare a set of representative detectors characterized by different recall and precision, and we test different inspection intervals ranging from 500ms to 10s. We evaluate the latency-cost trade-off achieved while considering two cases for the latency constraint associated with the service: a case where the service has a tight SLO latency (200ms) and a case where the SLO latency is less restrictive (500ms). To ensure that our results are not biased by the parameters used to instantiate the model, we also study the impact of varying some of these parameters, such as the anomaly rate, on the obtained results.

Our results show that achieving both a high precision and a high recall is not strictly necessary for balancing performance and cost. Depending on the inspection frequency used, focusing on obtaining high precision or high recall is sufficient for a close-to-optimal trade-off. With inspections at each second or less, prioritizing precision is most effective, while with lower inspection frequencies, emphasizing on recall becomes a better option.

To summarize, this paper makes the following contributions: (i) We present a model based on SRNs to study the performance and cost of a Cloud service monitored by an anomaly detector; (ii) Using this model, we run an extensive study of the main characteristics of performance anomaly detectors and their impact on the trade-off between latency and resource consumption achieved for the monitored service.


The rest of the paper is organized as follows. Section~\ref{sec:background} presents performance anomalies and anomaly detectors in Cloud services. We describe our model in Section~\ref{sec:model}. Section~\ref{sec:eval} details the results of our numerical evaluations. Finally, we discuss related work in Section~\ref{sec:rwork} and conclude in Section~\ref{sec:concl}.

\section{Performance Anomaly Detection in Cloud Services \label{sec:background}}

In this section, we define the performance anomalies considered throughout the paper. We also give details about the characteristics of anomaly detectors and how they can impact the cost and performance when used to observe Cloud services. Lastly, we define the research questions to be answered in the paper.

\subsection{Performance anomalies in the Cloud}


Modern Cloud applications or services can achieve high availability, fault isolation, and easier maintenance due to the distribution and replication of computation through multiple machines and datacenters~\cite{seemakhupt2023cloud}. Datacenters typically host a large number of services that share the available computing resources. Upon a major variation in workload, services can dynamically \textit{scale out} to use a greater share of these resources. However, assigning an appropriate amount of resources to each service is complex and bad decisions can severely impact their performance~\cite{seemakhupt2023cloud, iorgulescu2018perfiso, wydrowski2024load}.




Service-Level objectives (SLOs) define performance targets for a service. For metrics such as availability, latency, or throughput, SLOs define the quality of service users can expect. Performance anomalies can be defined as unexpected degradation in system behavior that negatively impact such metrics. Performance issues can arise from a variety of factors, including resource contention, bad configuration, software bugs, or transient hardware failures~\cite{zhou2018fault}. In this paper, we focus on performance anomalies that can be solved by scaling out, that is, creating more replicas of a service.


A typical example of performance anomaly is the case of \emph{antagonist} load, where activity from applications co-hosted on the same hardware resources as a service impact the performance of that service~\cite{wydrowski2024load}. Indeed, to optimize resource usage, it is common practice to co-locate latency-critical services with batch jobs~\cite{iorgulescu2018perfiso}. Jobs can be isolated using virtualization techniques with systems like Kubernetes~\cite{kubernetes} and resource oversubscription is usually done to ensure close to 100\% resource usage constantly~\cite{wydrowski2024load}. In this context, it was observed that the performance of latency-critical jobs can be severely impacted by the activity of co-hosted jobs. In some scenarios, an up-to 29x increase of the 99-percentile latency of a service was observed~\cite{iorgulescu2018perfiso}. In other cases, instances of services might even get killed because of resource exhaustion~\cite{zhou2018fault}.

Scaling-out a service experiencing a performance anomaly helps solving the problem as it reduces the load that each replica of the service should process. However, there is a risk of over-provisioning and consuming more resources than needed to handle the load. On the other hand, it is important to detect and react to performance anomalies early enough to limit the impact on end-users~\cite{ghosh2024fast}. Hence techniques to detect performance anomalies accurately and fast have been proposed, as we describe in the next section.

\subsection{Properties of anomaly detectors}


Performance anomaly detectors vary widely in their characteristics, most notably in recall, precision, and inspection frequency. On one hand, complex deep learning methods~\cite{li2021multivariate} can achieve high recall and precision, but often operate at low inspection frequencies because they require long processing times. On the other hand, heuristic approaches~\cite{ghosh2024fast} are easier to implement and have low processing times, but with lower recall and precision. Simpler machine learning methods~\cite{chow2022deeprest} are able to offer competitive recall and precision without greatly reducing inspection frequency.




In this work, we mainly analyze detectors in terms of precision and recall\footnote{Precision = \(\frac{TP}{TP + FP}\) and Recall = \(\frac{TP}{TP + FN}\), with TP, FP, and FN representing true positives, false positives, and false negatives, respectively.}.
Precision indicates the percentage of anomaly detections that are correct, while recall, the proportion of anomalies correctly detected. Some detectors allow tuning parameters to prioritize precision or recall. For example, using a tight detection threshold can reduce the total number of detections (true and false positives), consequently increasing precision but lowering recall. Deciding which metric to prioritize is challenging without considering the effect on the observed service.

Another important characteristic that we consider is inspection frequency, that is, how often the detector evaluates the state of the monitored service. A high inspection frequency allows anomalies to be detected faster, consequently reducing the time with degraded performance. However, high frequency is not always feasible. Certain detection methods take longer to process data or require slow data collection steps, which limits how often they can run. In addition, if the anomaly detector has low precision, performing frequent inspections can lead to a significant number of false detections and wasted resources due to unnecessary scaling actions.

\begin{figure}
\centering
\includegraphics[width=0.8\columnwidth]{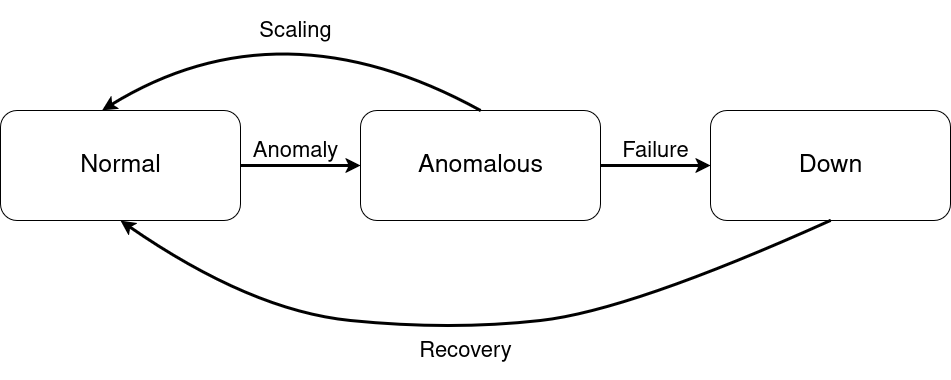}
  \caption{Service state transitions}
\label{fig:state}
\end{figure}

The state transition diagram of Figure~\ref{fig:state} illustrates the interactions between a performance anomaly detector and a monitored service. In the \texttt{normal} state, the service meets its SLO and delivers expected performance. When an anomaly occurs, such as interference from antagonist jobs, the service enters an \texttt{anomalous} state with increased latency. If the detector correctly identifies the anomaly, a scale-out action is triggered, restoring the service to the \texttt{normal} state. If the anomaly goes undetected, it could result in a crash. Then, the service remains in the \texttt{down} until restarting.





\subsection{Research questions}

In summary, the characteristics of an anomaly detector can influence the performance and cost of a service. Considering the latency as performance metric and resource consumption as cost metric, we aim to answer the following questions:


\begin{enumerate}
    \item What is the influence of the precision and the recall of the anomaly detector on the trade-off between performance and cost for a Cloud service?
    \item How does the inspection frequency influence this trade-off?
\end{enumerate}

\section{Modeling the System with SRNs\label{sec:model}}


This section outlines our model using Stochastic Reward Nets (SRNs) to analyze the performance and cost of a Cloud service with an anomaly detector. We evaluate service performance through average latency and asses cost based on resource consumption. We begin with an introduction to SRNs, followed by our assumptions about the Cloud service and the anomaly detector. Lastly, we present the SRN model and describe our evaluation methods for performance and cost.

\subsection{Introduction to SRNs}
Stochastic Reward Nets are an extension of Stochastic Petri Nets (SPNs) that offer a higher-level graph-based representation for the stochastic behaviors of dynamic systems \cite{trivedi2001probability}. A Petri net is represented as a directed bipartite graph consisting of two node types: places and transitions. Places are connected to transitions through directed arcs, which can either lead into or out of a transition. A marking of a Petri net is defined by the distribution of tokens across the places. Each unique marking represents a different state of the system. State changes occur when a transition fires, altering the marking of the Petri net. A transition becomes enabled when all its input places contain the required number of tokens. Upon firing, tokens are removed from the input places based on their multiplicity and are then added to the output places.

A Generalized Stochastic Petri Net (GSPN)~\cite{ajmone1984class} incorporates two kinds of transitions: timed and immediate. Timed transitions, depicted as blank rectangles, fire after a delay that follows an exponential distribution. In contrast, immediate transitions, represented by black lines, fire instantaneously, with no delay. Stochastic Reward Nets (SRNs) build upon GSPNs by adding guard functions and reward functions. Reward functions, specifically, assign values or rewards to tangible markings within the Petri net.

SRNs can be analyzed using software tools like SPNP \cite{ciardo1989spnp} and SHARPE \cite{sahner2012performance}. The state transitions defined within an SRN model can be mapped to an equivalent Continuous Time Markov Chain (CTMC). By solving the steady-state of the resulting CTMC, one can compute the expected reward rates specified by the reward functions. Further details on SRNs can be found in  \cite{trivedi2001probability, ciardo1993automated}.

\subsection{Modeled system}

Following the state transition diagram in Figure~\ref{fig:state}, we model a system composed of a Cloud service and an anomaly detector, where performance anomalies can occur. The rate in which anomalies occur is a parameter of the model. Here, the anomaly detector performs inspections periodically to determine the state of the service. This means that inspections are made both in the normal and anomalous states. If the anomaly is correctly identified, the service is scaled out and it returns to the normal state. If an anomaly is incorrectly detected in the normal state, the service is also scaled out and remains in the same state with extra resource usage. The inspection frequency is also a parameter of the model. In the case the anomaly is not detected quickly enough, then the service fails and crashes. The failure rate is equally a model parameter. Some further assumptions on the system behavior are made for our model:


\begin{enumerate}
    \item \textit{Default resource configuration}: We consider that a service is deployed with an initial resource configuration, which defines the default number of replicas of the service. 
    \item \textit{Scaling out after anomaly detection}: We consider that the anomaly detector is used only for detecting performance anomalies and triggering scaling out actions. In this case, a scaling out action after an anomaly is detected consists of deploying extra replicas of the service. 
    \item \textit{Down-scaling}: After the performance anomaly is handled and the service comes back to the normal state, we consider that the number of replicas of the service is down-scaled to the default value automatically after a certain time duration. Here, this duration is assumed to be long enough so that the original anomaly is completely handled.
    \item \textit{Resource configuration after service failure}: We assume that the service is restarted in a scaled out configuration after a failure, so that the potentially still existing anomaly can be handled. The number of replicas is also returned to the default value automatically after a certain time duration.
    \item \textit{Anomaly severity}: We do not model the case where a performance anomaly is resolved on its own. We assume that anomalies are sufficiently critical that they always lead to a service failure if not handled.
\end{enumerate}

\subsection{Modeling an anomaly detector for a Cloud service with SRNs}



In our model, we aim to represent how the different properties of a given anomaly detector influence the state transitions of a Cloud service. Since SRNs use transitions and places to represent state behaviors, the precision, recall, and inspection frequency of anomaly detection must be translated to rates and probabilities of transitions in the model. Similarly, other characteristics of service state changes in Figure~\ref{fig:state}, such as anomaly, failure, and recovery rates, are represented as transition rates. In the following, we start by defining the parameters of the model. Then, we describe our model in detail.

\subsubsection{Parameters of our model}

\begin{table}[h!]
 \small
	\caption{Description of SRN model variables}
 \label{table:rates}
\centering
 \begin{tabular}{c l} 
 \hline
 Variable & Description  \\ [0.5ex] 
 \hline
 $\lambda_{a}$ & Anomaly rate of a service in the normal state  \\ 
 $\lambda_{f}$ & Failure rate of a service in the anomalous state  \\
 $\mu$ & Recovery rate of a service in the down state  \\
 $\delta$ & Inspection rate of the anomaly detector \\
 $\sigma$ & Inference rate of the anomaly detector  \\
 $\gamma_{1}$ & Completion rate of scaling out action  \\ 
 $\gamma_{2}$ & Completion rate of down-scaling action  \\ 
 $p_{tp}$ & Probability of detecting a true positive  \\
 $p_{tn}$ & Probability of detecting a true negative \\
 $p_{fp}$ & Probability of detecting a false positive \\
 $p_{fn}$ & Probability of detecting a false negative \\ 
 $l_{n}$ & Average latency in the normal state \\
 $l_{a}$ & Average latency in the anomalous state \\
 $l_{d}$ & Average latency in the down state \\
 $r_{def}$ & Number of replicas in default configuration \\
 $r_{out}$ & Number of replicas in scaled out configuration \\
 [1ex] 
 \hline
 \end{tabular}

\end{table}

The rates and probabilities parameters for the anomaly detector and Cloud service are summarized in Table~\ref{table:rates}. Rate parameters refer to the number of times a transition is fired on average in a 1-hour period. The firing times of all transitions in the model follow an exponential distribution according to its specific firing rate parameter. In the case of immediate transitions, probability parameters are used instead, to directly define the likelihood of the transition being fired.   

The transition rates of the modeled Cloud service are given by: the anomaly rate ($\lambda_a$), the failure rate ($\lambda_f$), the recovery rate ($\mu$), the scaling out completion rate ($\gamma_1$) and the down-scaling completion rate ($\gamma_2$). The first three parameters ($\lambda_a$, $\lambda_f$, $\mu$) represent the transitions between normal, anomalous and down states. The scaling out and down-scaling completion rates refer to the average time needed for the change in a number of replicas to take effect.


The parameters referring to the anomaly detector are defined by: the inspection rate ($\delta$), the inference rate ($\sigma$), and the detection probabilities ($p_{tp}$, $p_{tn}$, $p_{fp}$, $p_{fn}$). The true positive ($p_{tp}$) and false negative ($p_{fn}$) probabilities represent the likelihood of detecting or not an anomaly in the anomalous state. Alternatively, the true negative ($p_{tn}$) and false positive ($p_{fp}$) probabilities are the chance of correctly or incorrectly identifying the normal state. Hence, the precision and recall of the anomaly detector can be defined through detection probabilities. Recall is simply the proportion of true anomalies detected. Precision is the quantity of true anomalies detected compared to the total number of detections, be them correct or incorrect. Such relations can be defined by equations:

\begin{equation}
    precision = \frac{A * p_{tp}}{A * p_{tp} + (1 - A) * p_{fp}} \\
\label{eq:prec}
\end{equation}
\begin{equation}
    recall = p_{tp} \\
\label{eq:rec}
\end{equation}

Here, $A$ is the percentage of anomaly observations in the data.
Lastly, the inspection rate ($\delta$) parameter refers to the inspection frequency of the detector, while inference rate ($\sigma$) represents its processing time. We only model scenarios where the processing time fits in the inspection frequency.

\subsubsection{The SRN model in detail}

\begin{figure}
\centering
\includegraphics[width=1.0\columnwidth]{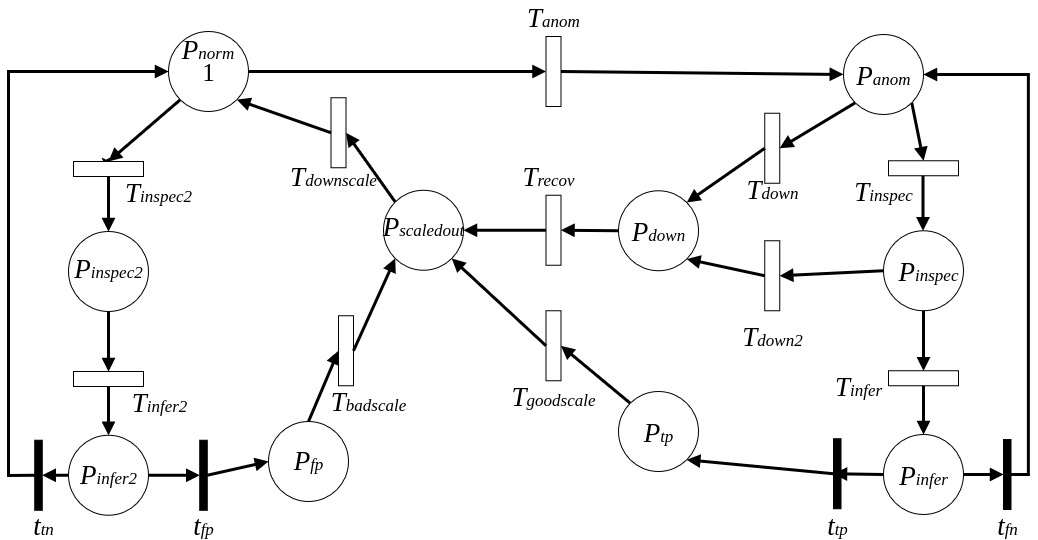}
  \caption{SRN model of service and anomaly detector.}
\label{fig:srnmodel}
\end{figure}

The graphical representation of our model with places and transitions is shown in Figure~\ref{fig:srnmodel}. In the initial state, a token is deposited in place \( P_{\text{norm}} \), representing the normal state of the service. Upon the occurrence of an anomaly, the transition \( T_{\text{anom}} \) is fired, which removes a token from \( P_{\text{norm}} \) and adds a new token to \( P_{\text{anom}} \). The anomaly rate \( \lambda_{a} \) is assigned to the \( T_{\text{anom}} \) transition.

The anomaly detector performs inspections regardless of the service state. Transitions \( T_{\text{inspec}} \) and \( T_{\text{inspec2}} \) are triggered when an inspection is initiated, both following inspection rate \( \delta \).

If an inspection is initiated while in the normal state, \( T_{\text{inspec2}} \) is fired, removing a token from \( P_{\text{norm}} \) and adding one to \( P_{\text{inspec2}} \). If the inspection starts in the anomalous state, \( T_{\text{inspec}} \) is fired instead, removing a token from \( P_{\text{anom}} \) and adding one to \( P_{\text{inspec}} \).

After the inspection begins, the anomaly detector requires some time to infer the current state. The inference time is modeled by transitions \( T_{\text{infer}} \) and \( T_{\text{infer2}} \), each assigned the inference rate \( \sigma \). When inference concludes in the normal state, \( T_{\text{infer2}} \) is fired, removing a token from \( P_{\text{inspec2}} \) and depositing another in \( P_{\text{infer2}} \). In the case of an anomalous state, \( T_{\text{infer}} \) is fired instead, removing a token from \( P_{\text{inspec}} \) and depositing a new one in \( P_{\text{infer}} \).

When the service is under the effect of a performance anomaly, the possible outcome of the anomaly detector are either a true positive (detection) or a false negative (fail to detect). These possibilities are represented by the immediate transitions \( t_{\text{tp}}\) and \( t_{\text{fn}}\) from place \( P_{\text{infer}} \), which are assigned probabilities \( p_{\text{tp}} \) and \( p_{\text{fn}} \), respectively. Probabilities are assigned so that $p_{tp}+p_{fn}=1$. If \( t_{\text{tp}}\) is fired instead of \( t_{\text{fn}}\), it represents the correct detection of an anomaly, in this case a token is removed from \( P_{\text{infer}} \) and a new one is added to \( P_{\text{tp}} \). After the anomaly detector identifies an anomaly, a scale out action is made to increase the capacity of the service and resolve the performance anomaly. The scaling out of replicas is represented by transition \( T_{\text{goodscale}} \), which takes one token from \( P_{\text{tp}} \) and adds another to \( P_{\text{scaledout}} \). The rate used for completion of a scaling action in transition \( T_{\text{goodscale}} \) is \( \gamma_{1} \). After scaling out, the service stays in a state of added replicas before automatically down-scaling to the default resource configuration. Down-scaling is then represented by transition \( T_{\text{downscale}} \) which removes a token from \( P_{\text{scaledout}} \) and adds a new one to \( P_{\text{norm}} \). The completion rate of the down-scaling process is represented by variable \( \gamma_{2} \). 

Alternatively, the firing of \( t_{\text{fn}}\) removes a token from \( P_{\text{infer}} \) and places a new one in \( P_{\text{anom}} \), representing no detection of the ongoing anomaly. If a performance anomaly is not detected, then the service continues in an anomalous state. When the anomaly remains undetected after one or more inspections, the service may fail and have to be restarted to recover to normal functioning state. This process is modeled through the transitions \( T_{\text{down}} \), \( T_{\text{down2}} \) and \( T_{\text{recov}} \). The transition \( T_{\text{down}} \) represents the case when the service fails before an inspection is made, while \( T_{\text{down2}} \) models the possibility of it failing in the period after the inspection has started and before the anomaly detector inference is completed. Both of these transitions are assigned the same rate \( \lambda_{f} \). The transition \( T_{\text{recov}} \) represents the recovery of a service after it has failed, which is assigned the rate \( \mu \). During recovery, the service is scaled out to handle the still existing anomaly. The firing of transition \( T_{\text{down} }\) adds a token in \( P_{\text{down}} \) while removing one from \( P_{\text{anom}} \). 

It is worth pointing out that we do not model a direct transition from $P_{infer}$ to $P_{down}$ because one of the transitions $p_{tp}$ or $p_{fn}$ is fired immediately when a token is deposited in $P_{infer}$. 

In the normal functioning state, the inspection process represented by the transitions \( T_{\text{inspec2}} \) and \( T_{\text{infer2}} \) follows the same behavior as in the anomalous state. However, the possible outcomes of the anomaly detector in this case are true negative (correct identification of normality) or a false positive (incorrect detection of anomaly). Here, the immediate transitions \( t_{\text{tn}}\) and \( t_{\text{fp}}\) from \( P_{\text{infer2}} \) represent these outcomes, which are assigned probabilities \( p_{tn} \) and \( p_{fp} \), respectively. Probabilities are assigned so that $p_{tn}+p_{fp}=1$. In the case of an incorrect detection, an unneeded scaling out action is represented by place \( P_{\text{fp}} \) and by transition \( T_{\text{badscale}} \). When \( t_{\text{fp}} \) is fired, a token is placed in \( P_{\text{fp}} \) and one is removed from \( P_{\text{infer2}} \). Upon the firing of \( T_{\text{badscale}} \), a token is then taken from \( P_{\text{fp}} \) and another put in \( P_{\text{scaledout}} \). Similarly to before, down-scaling is done by transition \( T_{\text{downscale}} \). Alternatively, upon the correct identification of normality, it does not require any actions. In this case, firing \( t_{\text{tn} }\) removes a token from \( P_{\text{infer2}} \) and adds one to \( P_{\text{norm}} \).

\subsection{Metrics for performance and cost analysis}

We evaluate the performance of the modeled service through the \textit{average latency} observed in each state. The normal, anomalous and down states have an associated latency measure. In our model, the normal state corresponds to the places $P_{\text{norm}}$, $P_{\text{inspec2}}$, $P_{\text{infer2}}$, $P_{\text{fp}},$ and $P_{\text{scaledout}}$; the anomalous state corresponds to $P_{\text{anom}}$, $P_{\text{inspec}}$, $P_{\text{infer}}$ and $P_{\text{tp}}$; the down state corresponds to $P_{\text{down}}$.

The cost of the service is evaluated by its \textit{resource consumption}. This metric is defined as the number of replicas of the service. Here, we consider that the service can be in two states: scaled-out or default. The scaled-out configuration refers to the place $P_{\text{scaledout}}$ where the service underwent a scaling-out action. The default resource configuration is represented by all other places. 

Table~\ref{table:rewards} shows the definition of the \textit{average latency} and the \textit{resource consumption} as reward functions. The notation $\#P_{\text{x}}$ represents the number of tokens in place $P_{\text{x}}$ and a test $\#P_{\text{x}} == 1$ checks if a token is ever in place $P_{\text{x}}$. By solving the SRN, we can estimate the probability that a token is in this place during the analysis.

The evaluation of service performance is estimated with the \textit{svlat} reward function, which computes its average latency:
\begin{equation}
    svlat = p(NormState)*l_n + p(AnomState)*l_a + p(DownState)*l_d
\end{equation}
Here, $p(\text{NormState})$ refers to the probability of the service being in the normal state during SRN analysis. The value $p(\text{AnomState})$ is the probability of the service being in the anomalous state. As for $p(\text{DownState})$, it represents the probability of the service being down. The parameters $l_n$, $l_a$, and $l_d$ correspond to the average latency of a service in the normal, anomalous, and down states, respectively. The values used for these parameters during our analysis are defined in Table~\ref{table:setvars}. Justifications for the chosen values are presented in Section~\ref{sec:eval_setup_env}. 

As for the evaluation of service cost, it is computed with the \textit{svcost} reward function. It estimates the mean number of replicas used by the service.
\begin{equation}
    svcost =  p(Default)*r_{def} + p(ScaledOut)*r_{out}
\end{equation}
Here, $p(\text{ScaledOut})$ and $p(\text{Default})$ denote the probabilities of the service being scaled out or running with the default number of replicas, respectively. We consider default resource consumption in $P_{\text{down}}$, since resources remain allocated on the host-machine even after a crash. This happens because the orchestrator does not free them immediately. The parameters $r_{def}$ and $r_{out}$ give the number of replicas in the default and scaled-out configurations, set as shown in Table~\ref{table:setvars} to represent a simple scaling-out scenario.

\begin{table}[h!]
\small
  \caption{Reward functions of the SRN model}
 \label{table:rewards}
\centering
 \begin{tabular}{l l l} 
 \hline
 Name & Measure & Function \\ [0.5ex] 
 \hline

 \textit{svlat} & \begin{tabular}{@{}l@{}} Average \\ latency \end{tabular}  & 
 \begin{tabular}{@{}l@{}} if  (( $\#P_{\text{anom}} == 1$ ) or ( $\#P_{\text{inspec}} == 1$ ) \\ or ( $\#P_{\text{infer}} == 1$ ) or  ( $\#P_{\text{tp}} == 1$ )) \\ then: $l_{a}$ \\ else if ( $\#P_{\text{down}} == 1$ ) \\ then: $l_{d}$ \\ else: $l_{n}$
    \end{tabular} \\
  \hline

 \textit{svcost} & \begin{tabular}{@{}l@{}} Resource \\ consumption \end{tabular}  & \begin{tabular}{@{}l@{}} if ( $\#P_{\text{scaledout}} == 1$ ) \\ then: $r_{out}$ \\ else: $r_{def}$  \end{tabular}\\
 [1ex]
 \hline
 \end{tabular}

\end{table}

\section{Evaluation \label{sec:eval}}
    This section presents the results of our numerical experiments of the proposed SRN model and discusses the impact of the characteristics of the anomaly detector on service performance and cost.

\subsection{Experimental setup}

Our SRN model is implemented and evaluated using SPNP~\cite{ciardo1989spnp}. For the numerical experiments, we fix the parameters related to the Cloud service while varying the parameters referring to the anomaly detector. A summary of the chosen values is displayed in Tables~\ref{table:setvars} and ~\ref{table:detecparams}. Further explanation about the choice of these values is presented below throughout the section. 






\subsubsection{System parameters\label{sec:eval_setup_env}}

\begin{table}[h!]
\small 
\caption{Values for SRN model variables}
 \label{table:setvars}
\centering
 \begin{tabular}{c l | c l } 
 \hline
 Variable & Value  & Variable & Value \\ [0.5ex] 
 \hline
 $\lambda_{a}$ & 6 ($h^{-1}$) & $l_{n}$ & 50 (ms) \\ 
 $\lambda_{f}$ & 60  ($h^{-1}$) & $l_{a}$ & 100 (ms) \\
 $\mu$ & 72 ($h^{-1}$) & $l_{d}$ & 25000 (ms) \\
 $\sigma$ & 36000 ($h^{-1}$) & $r_{def}$ &  1 (replica) \\
 $\gamma_{1}$ & 360 ($h^{-1}$) & $r_{out}$ & 2 (replicas) \\ 
 $\gamma_{2}$ & 60 ($h^{-1}$) \\ 
 [1ex] 
 \end{tabular}
\end{table}

\paragraph{Anomaly rate $\lambda_a$ and failure rate $\lambda_f$} 
The anomaly rate is not a metric that is openly available in the literature from production Cloud services. To estimate it, we searched available datasets on request traces of large Cloud services (Alibaba\footnote{\url{https://github.com/alibaba/clusterdata/tree/master/cluster-trace-microservices-v2022.}} and Azure\footnote{\url{https://github.com/Azure/AzurePublicDataset/blob/master/AzureFunctionsDataset2019.md}}) for moments of high intensity which could be expected to cause performance anomalies. As detailed in Section~\ref{sec:background}, we assume that moments with a significant increase in workload intensity can be the cause of performance interferences on co-located services. By analyzing 1-hour extracts of different services from these datasets, we obtain an estimate rate of workload intensity peaks per hour in these services. The resulting workload peak rates obtained fell in the range of 1 to 20 per hour. Considering that not all moments of high intensity necessarily generate performance anomalies, we set a default anomaly rate $\lambda_a = 6$ anomalies per hour. We test different values of $\lambda_a$ later in Section~\ref{sec:eval_sens}.


The failure rate $\lambda_f$ of a service in the anomalous state is also not easily found in the literature. For this, we conducted local anomaly injection experiments on the \textit{SocialNetwork} and \textit{HotelReservation} applications from the DeathStarBench benchmark suite\footnote{\url{https://github.com/delimitrou/DeathStarBench}}, deployed in a Kubernetes cluster~\cite{kubernetes}. The anomaly injection experiments consisted of generating artificial memory interferences in certain services. We observed gradual performance degradation until the service would crash. From the results, we observed that the time between the start of the anomaly injection and the service crash would fall in a range from 5 to 120 seconds. The time to crash varied depending on how stressed the service was at the time of anomaly injection and depending on specific characteristics of the service. Thus, we assume a general case where the average time to crash is 60 seconds for our SRN model evaluation, which is represented by a failure rate $\lambda_f = 60$ crashes per hour. We also test different values of $\lambda_f$ in Section~\ref{sec:eval_sens}.

\paragraph{Recovery rate $\mu$ and scaling completion rates $\gamma_1,\gamma_2$ }
To set parameters referring to Kubernetes behavior, such as the recovery rate $\mu$ and the scaling-out completion rates $\gamma_1$, we refered to existing work~\cite{vayghan2019kubernetes} and Kubernetes documentation. Upon service failure, by default Kubernetes considers a 30-second wait time for a service to be gracefully terminated and an initial 10-second backoff delay to restart a service after it is terminated. Assuming that services might need to execute startup routines before it is ready for handling requests (e.g. database synchronization routine), we consider a mean time of 10 seconds for a service to be ready after a restart. Hence, the average recovery time of a service after a failure is estimated to 50 seconds, which can be represented by a recovery rate $\mu = 72h^{-1}$. In the case of the deployment of a new replica, that is, scaling out a service, we only consider the 10 second interval for it to be ready, which gives an up-scaling completion rate $\gamma_1 = 360h^{-1}$. 

The down-scaling rate $\gamma_2$ represents the time delay before a service is automatically down-scaled following a scaling-out action. We assume an average period of 60 seconds with increased resource consumption for the anomaly to be considered fully resolved. This corresponds to a down-scaling rate of $\gamma_2 = 60h^{-1}$.

\paragraph{Average latencies $l_n,l_a,l_d$ and number of replicas $r_{def},r_{out}$}
The system variables used by the \textit{svlat} reward function (Table~\ref{table:rewards}), that is the average latencies $l_{n}$, $l_{a}$, and $l_{d}$, are mostly based on observations from real production systems at Google and Uber. Notably, Google~\cite{seemakhupt2023cloud,wydrowski2024load} observations indicates that the median latency of a service during normal execution remains in a range between 10 and 100 ms. Hence, in our model we assume an average latency of $l_n=50$ms in the normal state of the service. Complementary observations from Uber~\cite{lee2024tale} and Google~\cite{seemakhupt2023cloud} services show that latency can increase by up to 100\% or more in situations analogous to our definition of performance anomalies. Based on this, we chose to consider an average latency $l_a=100$ms when a service is under the influence of a performance anomaly.

For the average latency of a crashed service ($l_d$), we assume that request delays range from the full recovery time to almost zero. Specifically, a request arriving exactly at the moment of the crash would wait the maximum time (50 seconds), while a request arriving just before the service restarts would wait the minimum (close to 0 seconds). Taking the mean of this range, we set $l_d = 25000$ms.

The number of replicas used by the \textit{svcost} reward function $r_{def}$ and $r_{out}$ are set to represent the simplest scaling-out scenario. That is, $r_{def}=1$ and $r_{out}=2$, which represents the service being scaled-out from 1 to 2 replicas. We believe this is sufficient to measure the impact of detectors on service resource consumption cost.


\subsubsection{Anomaly detector parameters}

\paragraph{True positive, false positive, true negative, and false negative probabilities}
To analyze the impact of the precision and recall of an anomaly detector on the performance and cost of a service, we compare different representative detectors. These detectors are represented by a combination of recall and precision values. All detector flavors tested are defined in Table~\ref{table:detecmetrics}. The \textit{Superior} detector reflects a detector with high recall and precision. The \textit{GreatPrec} and \textit{GoodPrec} detectors demonstrate high precision and average recall, with \textit{GoodPrec} having a slightly lower precision than \textit{GreatPrec}; both are representative of detectors that generate few false positives at the expense of also generating few true positives. Inversely, the \textit{GreatRec} and \textit{GoodRec} detectors show high recall at the expense of lower precision. The small difference in precision between \textit{GreatRec} and \textit{GoodRec} is clarified later in the section. The \textit{Heuristic} detector represents the case of simpler detection techniques that provide good recall, but lower precision than previous detectors. Lastly, the \textit{Random} detector serves as a baseline that has a 50\% probability of detecting an anomaly in any of the states, that is, $p_{tp} = p_{fp} = p_{tn} = p_{fn} = 0.5$.

\begin{table}[h!]
\small
 \caption{Precision and recall of tested detectors}
 \label{table:detecmetrics}
\centering
 \begin{tabular}{l | c c || l | c c} 
 \hline
 Detector & Recall & Precision & Detector & Recall & Precision\\ [0.5ex] 
 \hline
 \textit{Superior} & 0.95 & 0.97 & \textit{Random} & 0.5 & 0.04\\ 
 \textit{GreatPrec} & 0.5 & 0.95 & \textit{GreatRec} & 0.95 & 0.79 \\
 \textit{GoodPrec} & 0.5 & 0.8 & \textit{GoodRec} & 0.70 & 0.74\\ 
 \textit{Heuristic} & 0.9 & 0.42 & & & \\
 \end{tabular}
\end{table}

The selected precision and recall values for the representative detectors were based by existing anomaly detection approaches in the literature~\cite{chow2022deeprest, gan2021sage, li2021multivariate, christofidi2023machine, chatfield2019analysis}. To validate these choices, we conducted local experiments using various detector implementations on a dataset from the same DeathStarBench environment described in Section~\ref{sec:eval_setup_env}, which contained approximately 4\% anomalous observations. The results were consistent with the precision and recall values reported in the literature.

As explained in Section~\ref{sec:model}, our model uses detection probabilities $p_{tp}$, $p_{fp}$, $p_{tn}$, and $p_{fn}$ to model the possible outcomes of correct and incorrect anomaly detection. Hence, the precision and recall values of the tested detectors are translated into probabilities using Equations~\ref{eq:prec} and ~\ref{eq:rec} assuming an anomaly percentage $A=4\%$. Table~\ref{table:detecparams} summarizes the true positive and false positive probabilities for each detector, while the true negative and false negative probabilities are calculated from equations $p_{tn} = 1 - p_{fp}$ and $p_{fn} = 1 - p_{tp}$, respectively. It is worth pointing out that, while recall and true positive probability have a direct correlation, precision correlates to both true positive and false positive probabilities. This explains the difference in precision of \textit{GreatRec} and \textit{GoodRec}, despite equal false positive probability.

\begin{table}[h!]
\small 
\caption{True positive and false positive probabilities of tested detectors}
 \label{table:detecparams}
\centering
 \begin{tabular}{l | c c || l | c c } 
 \hline
 Detector & $p_{tp}$ & $p_{fp}$ & Detector & $p_{tp}$ & $p_{fp}$ \\ [0.5ex] 
 \hline
 \textit{Superior} & 0.95 & 0.001 & \textit{Random} & 0.5 & 0.5\\ 
 \textit{GreatPrec} & 0.5 & 0.001 & \textit{GreatRec} & 0.95 & 0.01 \\ 
 \textit{GoodPrec} & 0.5 &  0.005 & \textit{GoodRec} & 0.7 &  0.01 \\ 
 \textit{Heuristic} & 0.9 & 0.05 & & &\\
 
 \end{tabular}
\end{table}

\paragraph{Inspection rate $\delta$ and inference rate $\sigma$}

To evaluate the impact of the inspection frequency, we vary two model parameters: the inspection rate $\delta$, or how often the detector initiates an inspection, and the inference rate $\sigma$, or how long the detector takes to produce a result once inspection starts. Notably, we test $\delta$ using values from the set $\{18000h^{-1}, 3600h^{-1}, 720h^{-1}, 360h^{-1}\}$, which correspond to inspection intervals of $\{0.5s, 1s, 5s, 10s\}$.

For the inference rate $\sigma$, we varied it between intervals of $\{0.1s,$ $0.5s,$ $1s,$ $2s,$ $5s\}$ and found that it had little impact on our evaluation results. Hence in following evaluations, inference rate is fixed at $\sigma = 36000$, corresponding to an inference taking 0.1 seconds.


\subsection{Performance and cost analysis of detector precision, recall and inspection frequency}

\begin{figure}[t]
    \centering
    \small
    \begin{tikzpicture}
        \begin{axis}[
            hide axis,
            xmin=0, xmax=1,
            ymin=0, ymax=1,
            legend columns=4,
            legend style={
                at={(0.5,-0.1)},
                anchor=north,
                draw=none,
                /tikz/every even column/.append style={column sep=0.2cm}
            }
        ]
        \addlegendimage{blue, thick, solid, mark=asterisk}
        \addlegendentry{Superior}
        \addlegendimage{orange, thick, dashdotted , mark=diamond*}
        \addlegendentry{GreatPrec}
        \addlegendimage{green!70!black, thick, dashed , mark=x}
        \addlegendentry{GoodPrec}
        \addlegendimage{red, thick, dashdotted , mark=triangle*}
        \addlegendentry{GreatRec}
        \addlegendimage{purple, thick, dashed , mark=+}
        \addlegendentry{GoodRec}
        \addlegendimage{magenta, thick, solid, mark=o}
        \addlegendentry{Heuristic}
        \addlegendimage{brown, thick, solid, mark=*}
        \addlegendentry{Random}
        \end{axis}
    \end{tikzpicture}

   \makebox[\columnwidth][c]{%
        \begin{subfigure}[b]{0.48\columnwidth}
            \centering
            \begin{tikzpicture}[baseline=(current bounding box.north),font=\small]
\begin{axis}[
    xlabel={Inspection Interval (s)},
    ylabel={Average Latency (s)},
    xlabel style={at={(axis description cs:0.5,-0.1)}, anchor=north},
    ylabel style={at={(axis description cs:-0.1,0.5)}, anchor=south},
    width=5cm,
    height=4cm,
    grid=major
]

\addplot [blue, thick, solid, mark=asterisk] table [x index=7, y index=0, col sep=comma] {plots_tex/sensitivity_Latency.csv};
\addplot [orange, thick, dashdotted , mark=diamond*] table [x index=7, y index=1, col sep=comma] {plots_tex/sensitivity_Latency.csv};
\addplot [green!70!black, thick, dashed , mark=x] table [x index=7, y index=2, col sep=comma] {plots_tex/sensitivity_Latency.csv};
\addplot [red, thick, dashdotted , mark=triangle*] table [x index=7, y index=3, col sep=comma] {plots_tex/sensitivity_Latency.csv};
\addplot [purple, thick, dashed , mark=+] table [x index=7, y index=4, col sep=comma] {plots_tex/sensitivity_Latency.csv};
\addplot [magenta, thick, solid, mark=o] table [x index=7, y index=5, col sep=comma] {plots_tex/sensitivity_Latency.csv};
\addplot [brown, thick, solid, mark=*] table [x index=7, y index=6, col sep=comma] {plots_tex/sensitivity_Latency.csv};

\end{axis}
\end{tikzpicture}
            \caption{Average latency.}
            \label{fig:latency}
        \end{subfigure}
        \hfill
        \begin{subfigure}[b]{0.48\columnwidth}
            \centering
            \begin{tikzpicture}[baseline=(current bounding box.north),font=\small]
\begin{axis}[
    xlabel={Inspection Interval (s)},
    ylabel={Resource Cons. (nb. replicas)},
    xlabel style={at={(axis description cs:0.5,-0.1)}, anchor=north},
    ylabel style={at={(axis description cs:-0.1,0.5)}, anchor=south},
    width=5cm,
    height=4cm,
    grid=major
]

\addplot [blue, thick, solid, mark=asterisk] table [x index=7, y index=0, col sep=comma] {plots_tex/sensitivity_ScalingCost.csv};
\addplot [orange, thick, dashdotted , mark=diamond*] table [x index=7, y index=1, col sep=comma] {plots_tex/sensitivity_ScalingCost.csv};
\addplot [green!70!black, thick, dashed , mark=x] table [x index=7, y index=2, col sep=comma] {plots_tex/sensitivity_ScalingCost.csv};
\addplot [red, thick, dashdotted , mark=triangle*] table [x index=7, y index=3, col sep=comma] {plots_tex/sensitivity_ScalingCost.csv};
\addplot [purple, thick, dashed , mark=+] table [x index=7, y index=4, col sep=comma] {plots_tex/sensitivity_ScalingCost.csv};
\addplot [magenta, thick, solid, mark=o] table [x index=7, y index=5, col sep=comma] {plots_tex/sensitivity_ScalingCost.csv};
\addplot [brown, thick, solid, mark=*] table [x index=7, y index=6, col sep=comma] {plots_tex/sensitivity_ScalingCost.csv};

\end{axis}
\end{tikzpicture}
            \caption{Resource consumption.}
            \label{fig:cost}
        \end{subfigure}
    }

    \caption{Analysis of average latency and resource consumption with varying detector configurations.}
    \label{fig:sensitivity}
\end{figure}
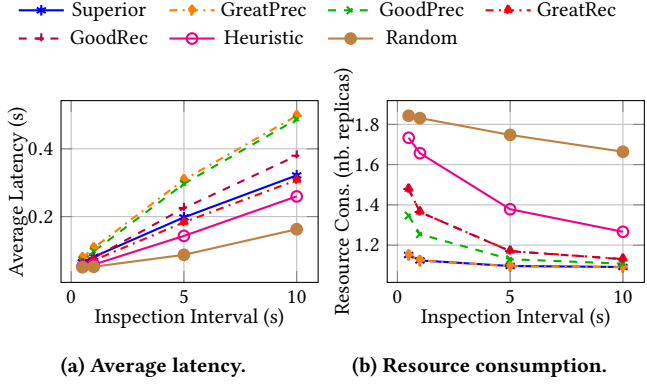

In this evaluation, we test different detector configurations to study the impact of an anomaly detector on the performance and cost of a service. To this end, we conduct numerical analyses with our model, applying the detector parameters listed in Table~\ref{table:detecparams} while varying the inspection interval. The results are presented in Figure~\ref{fig:sensitivity}. 

When analyzing the latency results in Figure~\ref{fig:latency}, we can observe that the most optimized detector (\textit{Superior}) does not necessarily make the service achieve the best average latency. Instead, simpler approaches such as \textit{Random} and \textit{Heuristic} present a lower average latency for the service in all tested inspection intervals. Notably, \textit{Random}, which has a worse recall and precision than \textit{Heuristic} and all other detectors, was able to maintain the best performance. Additionally, \textit{GreatRec}, which has the same recall and worse precision than \textit{Superior}, allowed for a generally better average latency than \textit{Superior}. Such results indicate that precision can actually be harmful to performance. This occurs because low-precision detectors, with a high false positive probability $p_{fp}$, often misclassify the normal state as anomalous and trigger unnecessary scaling out. In our SRN model, this increases the total time spent in place $P_{scaledout}$, which in turn reduces time spent in $P_{norm}$ and the chance of $T_{anom}$ transition firing. In real-world systems, this corresponds to the effect of having over-provisioned resources, thus lowering the likelihood of experiencing performance anomalies. 

Still in Figure~\ref{fig:latency}, we can observe from the comparison between \textit{GreatRec} and \textit{GoodRec} that recall remains an important factor to reduce the average latency of the service. Despite \textit{GoodRec} having a similar precision to \textit{GreatRec}, the higher recall of the latter allows for a gain in performance. This indicates that increasing the chance of detecting anomalies, that is, the true positive probability $p_{tp}$, is good for performance.



Regarding the cost analysis in Figure~\ref{fig:cost}, we can observe that it is highly influenced by the precision, that is, the false positive probability $p_{fp}$. Detectors with the highest precision such as \textit{Superior} and \textit{GreatPrec} consistently yield the lowest resource consumption across all inspection intervals. Furthermore, detectors with the same precision but different recall, such as \textit{Superior} and \textit{GreatPrec} or \textit{GreatRec} and \textit{GoodRec}, present identical costs. This shows that recall has little effect on resource consumption, and that the latter is mostly influenced by incorrect detections in the normal state.

As evident in both Figures~\ref{fig:latency} and~\ref{fig:cost}, the inspection interval has a major impact on both average  latency and resource consumption. In the case of average latency, short inspection intervals are key to achieving low latency, as all the detectors tested are capable of keeping the average latency below $0.1s$ for the inspection intervals of $0.5s$ and $1s$. Notably, a short inspection interval is important to increase the probability of an anomaly being detected and handled before a service crash occurs. Additionally, longer inspection intervals cause the service to stay in an anomaly state longer, where the latency is degraded. However, longer inspection intervals also correspond to less frequent inspections, which reduces the likelihood of false positives, and thus unneeded scaling. Hence, larger intervals reduce the cost, as is evident in Figure~\ref{fig:cost}.

In addition to detector flavors, we also evaluate a baseline "no detection" scenario using a simplified SRN model without any anomaly detection. The corresponding curves of this "no detection" scenario are not included in the presented plots since they would render other curves unreadable. The "no detection" scenario results in an average latency of $1.67$ seconds. This demonstrates the benefits of implementing a performance anomaly detection approach to maintain good service performance. In the same scenario, the service presents a cost of around 1.08 replicas, as the increase in the number of replicas only happens when the service is restarted after a crash. This indicates that the resource consumption achieved by \textit{Superior} and \textit{GreatPrec} with any inspection interval, and by \textit{GoodPrec}, \textit{GreatRec} and \textit{GoodRec} with higher intervals, is very close to minimal while maintaining much lower average latency values.

\subsection{Trade-off between performance and cost \label{sec:latcostscore}}


Previous results only show how parameters of a detector affect performance or cost independently, making it difficult to determine which detector best balances both. Hence, we use a weighted average to define a trade-off score between resource consumption and average latency. We define the \textbf{latency-cost score} $LC$:

\begin{equation}
    LC = \frac{w_l*l_{norm} + w_c*c_{norm}}{w_l + w_c} 
    \label{eq:lc}
\end{equation}

where $w_l$ and $w_c$ represent the associated weights for latency and cost, respectively. In our evaluations, we fix the weights $w_l = w_c = 1$. The variables $l_{norm}$ and $c_{norm}$ refer to latency and cost values scaled  using \textit{min-max} normalization \footnote{Given a variable $x$, the equation for min-max normalization is: $x_{norm} = \frac{x - x_{min}}{x_{max} - x_{min}}$}. The set of detector parameters that results in the lowest latency-cost score is considered to offer the best trade-off overall. 

For cost normalization $c_{norm}$, we consider that its maximum possible value is an over-provisioning case with resource consumption at 2 replicas, which gives $c_{max} = 2$ replicas. In addition, the minimum achievable cost is considered to be the case when no false positives happen and the number of replicas is only increased to handle an anomaly. In this case, the time the service spends with an additional replica depends on the average down-scaling time ($\gamma_2^{-1}$), and on the average number of anomalies over time ($\lambda_a$). By multiplying both of these values, we get the proportion of additional replicas per hour. By adding it to the base replica of a service, we obtain its minimum resource consumption, or $c_{min} = \lambda_a * \gamma_2^{-1} + 1$ replicas on average over time.

Regarding latency normalization, we set the minimum latency $l_{min} = 50ms$ to reflect the minimum obtainable latency in our model. Then, we consider that the maximum latency would be defined by the service SLO. Thus, we perform tests on example scenarios with different latency SLOs, $l_{max} = 500ms$ and $l_{max} = 200ms$. In practice,
reducing $l_{max}$ increases the value of $l_{norm}$, consequently raising the relative importance of latency in Equation~\ref{eq:lc}. For instance, a latency measure of $l=100ms$ gives $l_{norm}=0.11$ with $l_{max}=500ms$, where the same latency $l$ using $l_{max}=200ms$ would give  $l_{norm}=0.33$.

\begin{figure}[t]
    \small
    \centering
    \begin{tikzpicture}
        \begin{axis}[
            hide axis,
            xmin=0, xmax=1,
            ymin=0, ymax=1,
            legend columns=4,
            legend style={
                at={(0.5,-0.1)},
                anchor=north,
                draw=none,
                /tikz/every even column/.append style={column sep=0.2cm}
            }
        ]
        \addlegendimage{blue, thick, solid, mark=asterisk}
        \addlegendentry{Superior}
        \addlegendimage{orange, thick, dashdotted , mark=diamond*}
        \addlegendentry{GreatPrec}
        \addlegendimage{green!70!black, thick, dashed , mark=x}
        \addlegendentry{GoodPrec}
        \addlegendimage{red, thick, dashdotted , mark=triangle*}
        \addlegendentry{GreatRec}
        \addlegendimage{purple, thick, dashed , mark=+}
        \addlegendentry{GoodRec}
        \addlegendimage{magenta, thick, solid, mark=o}
        \addlegendentry{Heuristic}
        \addlegendimage{brown, thick, solid, mark=*}
        \addlegendentry{Random}
        \end{axis}
    \end{tikzpicture}

    \makebox[\columnwidth][c]{%
	\begin{subfigure}[b]{0.48\columnwidth}
            \centering
            \begin{tikzpicture}[baseline=(current bounding box.north),font=\small]
\begin{axis}[
    xlabel={Inspection Interval (s)},
    ylabel={Latency-Cost Score},
    xlabel style={at={(axis description cs:0.5,-0.1)}, anchor=north},
    ylabel style={at={(axis description cs:-0.1,0.5)}, anchor=south},
    width=5cm,
    height=4cm,
    grid=major
]

\addplot [blue, thick, solid, mark=asterisk] table [x index=7, y index=0, col sep=comma] {plots_tex/sensitivity_LatCostScore.csv};
\addplot [orange, thick, dashdotted , mark=diamond*] table [x index=7, y index=1, col sep=comma] {plots_tex/sensitivity_LatCostScore.csv};
\addplot [green!70!black, thick, dashed , mark=x] table [x index=7, y index=2, col sep=comma] {plots_tex/sensitivity_LatCostScore.csv};
\addplot [red, thick, dashdotted , mark=triangle*] table [x index=7, y index=3, col sep=comma] {plots_tex/sensitivity_LatCostScore.csv};
\addplot [purple, thick, dashed , mark=+] table [x index=7, y index=4, col sep=comma] {plots_tex/sensitivity_LatCostScore.csv};
\addplot [magenta, thick, solid, mark=o] table [x index=7, y index=5, col sep=comma] {plots_tex/sensitivity_LatCostScore.csv};
\addplot [brown, thick, solid, mark=*] table [x index=7, y index=6, col sep=comma] {plots_tex/sensitivity_LatCostScore.csv};

\end{axis}
\end{tikzpicture}
            \caption{Latency SLO of $l_{max}=500ms$.}
            \label{fig:latcost500}
        \end{subfigure}
        \begin{subfigure}[b]{0.48\columnwidth}
            \centering
            \begin{tikzpicture}[baseline=(current bounding box.north),font=\small]
\begin{axis}[
    xlabel={Inspection Interval (s)},
    xlabel style={at={(axis description cs:0.5,-0.1)}, anchor=north},
    width=5cm,
    height=4cm,
    grid=major
]

\addplot [blue, thick, solid, mark=asterisk] table [x index=7, y index=0, col sep=comma] {plots_tex/sensitivity_LatCostScore_200ms.csv};
\addplot [orange, thick, dashdotted , mark=diamond*] table [x index=7, y index=1, col sep=comma] {plots_tex/sensitivity_LatCostScore_200ms.csv};
\addplot [green!70!black, thick, dashed , mark=x] table [x index=7, y index=2, col sep=comma] {plots_tex/sensitivity_LatCostScore_200ms.csv};
\addplot [red, thick, dashdotted , mark=triangle*] table [x index=7, y index=3, col sep=comma] {plots_tex/sensitivity_LatCostScore_200ms.csv};
\addplot [purple, thick, dashed , mark=+] table [x index=7, y index=4, col sep=comma] {plots_tex/sensitivity_LatCostScore_200ms.csv};
\addplot [magenta, thick, solid, mark=o] table [x index=7, y index=5, col sep=comma] {plots_tex/sensitivity_LatCostScore_200ms.csv};
\addplot [brown, thick, solid, mark=*] table [x index=7, y index=6, col sep=comma] {plots_tex/sensitivity_LatCostScore_200ms.csv};

\end{axis}
\end{tikzpicture}
            \caption{Latency SLO of $l_{max}=200ms$.}
            \label{fig:latcost200}
        \end{subfigure}
    }

    \caption{Latency-cost score for different detectors on two distinct latency SLO values.}
    \label{fig:latcostmaxlat}
\end{figure}
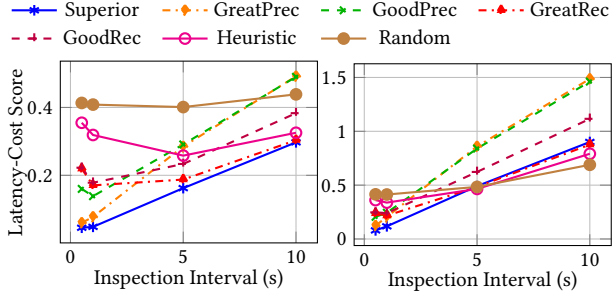

Figure~\ref{fig:latcostmaxlat} displays the latency-cost scores for the detectors defined in Table~\ref{table:detecparams} with SRN parameters from Table~\ref{table:setvars}. In a scenario with a desired latency SLO of $l_{max} = 500ms$ (Fig.~\ref{fig:latcost500}), it can be observed that for inspection intervals of $0.5s$ and $1s$, the latency-cost score presents a pattern similar to the results for the resource consumption (Fig.~\ref{fig:cost}). In other words, detectors with high precision achieve the lowest score and the best trade-off. This similarity can be explained as follows: for small inspection intervals, all detectors achieve similar latency values (Fig.~\ref{fig:latency}). Consequently, the score becomes mainly dictated by the resource consumption cost.

Interestingly, the best trade-off between average latency and resource consumption cost is achieved by both \textit{Superior} and \textit{GreatPrec} detectors, despite \textit{GreatPrec} having a much lower recall than \textit{Superior}. This suggests that, when choosing detectors for frequent inspections, precision is more critical than recall. Moreover, a detector like \textit{GreatPrec} requires considerably less optimization efforts to obtain in practice than one highly optimized as \textit{Superior}.

For inspection intervals of $5s$ and $10s$, the latency-cost score increases significantly for detectors \textit{Superior}, \textit{GreatPrec}, \textit{GoodPrec}, and \textit{GoodRec}. The high latency values (Fig.~\ref{fig:latency}) become more prevalent than the low cost achieved (Fig.~\ref{fig:cost}). In higher inspection intervals, we can also note that \textit{GreatRec} takes the place of \textit{GreatPrec} and becomes the detector with the closest score to \textit{Superior}. This indicates that recall becomes a more relevant optimization goal as the inspection interval increases. Furthermore, \textit{Heuristic} becomes one of the best options at an inspection interval of $10s$, which would be simpler to implement in practice than \textit{Superior} or \textit{GreatRec}.

In a more constrained latency SLO of $l_{max} = 200ms$, as shown in Figure~\ref{fig:latcost200}, the results with inspection intervals of $0.5s$ and $1s$ remain similar to the case with $l_{max} = 500ms$, as the average latency achieved by all detectors is relatively low. For bigger inspection intervals, the constraint $l_{max}$ increases the importance of latency in the latency-cost score, consequently making simpler detectors such as \textit{Heuristic} and \textit{Random} present a better trade-off between latency and cost than more complex approaches as \textit{Superior}, \textit{GreatPrec}, \textit{GoodPrec}, \textit{GreatRec} and \textit{GoodRec}. In summary, similar to the case with $l_{max} = 500ms$, high precision enables a better trade-off at smaller inspection intervals, while high recall becomes more beneficial at bigger inspection intervals.

Ultimately, the decision to prioritize recall or precision in an anomaly detector depends on the practical constraints of the deployment environment. For example, some detectors may only operate at longer inspection intervals due to processing or data collection overhead. While a detector with both high precision and recall, such as \textit{Superior}, generally offered the best trade-off between performance and cost across most scenarios, our results show that alternative detectors can achieve similar trade-offs when recall or precision is prioritized appropriately. In scenarios with inspection intervals smaller than or equal to $1\,\text{s}$, prioritizing precision yields great results, whereas for intervals greater than or equal to $5\,\text{s}$, emphasizing on recall proves more effective. In latency-critical service with tight SLOs, detectors with high recall and lower precision, despite higher resource consumption or potential over-provisioning, become a reasonable option.



\subsection{Impact of anomaly rate and time to crash\label{sec:eval_sens}}

The analyses in previous sections were made assuming a specific anomaly rate with on average 6 anomalies per hour and a mean time to crash in the anomalous state of 60 seconds. Different values for these rates may change the impact of the detectors on the average latency and resource consumption. In this section, we perform a sensitivity analysis of rates $\lambda_a$ and $\lambda_f$ to evaluate how different anomaly scenarios impact the obtained latency-cost score.

\begin{figure}[t]
   \small 
   \centering
    \begin{tikzpicture}
        \begin{axis}[
            hide axis,
            xmin=0, xmax=1,
            ymin=0, ymax=1,
            legend columns=4,
            legend style={
                at={(0.5,-0.1)},
                anchor=north,
                draw=none,
                /tikz/every even column/.append style={column sep=0.2cm}
            }
        ]
        \addlegendimage{blue, thick, solid, mark=asterisk}
        \addlegendentry{Superior}
        \addlegendimage{orange, thick, dashdotted , mark=diamond*}
        \addlegendentry{GreatPrec}
        \addlegendimage{green!70!black, thick, dashed , mark=x}
        \addlegendentry{GoodPrec}
        \addlegendimage{red, thick, dashdotted , mark=triangle*}
        \addlegendentry{GreatRec}
        \addlegendimage{purple, thick, dashed , mark=+}
        \addlegendentry{GoodRec}
        \addlegendimage{magenta, thick, solid, mark=o}
        \addlegendentry{Heuristic}
        \addlegendimage{brown, thick, solid, mark=*}
        \addlegendentry{Random}
        \end{axis}
    \end{tikzpicture}

    \makebox[\columnwidth][c]{%
        \begin{subfigure}[b]{0.48\columnwidth}
            \centering
            \begin{tikzpicture}[baseline=(current bounding box.north),font=\small]
\begin{axis}[
    xlabel={Anomaly Rate ($h^{-1}$)},
    ylabel={Latency-Cost Score},
    xlabel style={at={(axis description cs:0.5,-0.1)}, anchor=north},
    ylabel style={at={(axis description cs:-0.1,0.5)}, anchor=south},
    width=5cm,
    height=4cm,
    grid=major
]

\addplot [blue, thick, solid, mark=asterisk] table [x index=7, y index=0, col sep=comma] {plots_tex/sensitivity_LatCostScore_AnomalyRate.csv};
\addplot [orange, thick, dashdotted , mark=diamond*] table [x index=7, y index=1, col sep=comma] {plots_tex/sensitivity_LatCostScore_AnomalyRate.csv};
\addplot [green!70!black, thick, dashed , mark=x] table [x index=7, y index=2, col sep=comma] {plots_tex/sensitivity_LatCostScore_AnomalyRate.csv};
\addplot [red, thick, dashdotted , mark=triangle*] table [x index=7, y index=3, col sep=comma] {plots_tex/sensitivity_LatCostScore_AnomalyRate.csv};
\addplot [purple, thick, dashed , mark=+] table [x index=7, y index=4, col sep=comma] {plots_tex/sensitivity_LatCostScore_AnomalyRate.csv};
\addplot [magenta, thick, solid, mark=o] table [x index=7, y index=5, col sep=comma] {plots_tex/sensitivity_LatCostScore_AnomalyRate.csv};
\addplot [brown, thick, solid, mark=*] table [x index=7, y index=6, col sep=comma] {plots_tex/sensitivity_LatCostScore_AnomalyRate.csv};

\end{axis}
\end{tikzpicture}
            \caption{Anomaly rate}
            \label{fig:latcostanom}
        \end{subfigure}
        \begin{subfigure}[b]{0.48\columnwidth}
            \centering
            \begin{tikzpicture}[baseline=(current bounding box.north),font=\small]
\begin{axis}[
    xlabel={Time to crash (s)},
    xlabel style={at={(axis description cs:0.5,-0.1)}, anchor=north},
    width=5cm,
    height=4cm,
    grid=major
]

\addplot [blue, thick, solid, mark=asterisk] table [x index=7, y index=0, col sep=comma] {plots_tex/sensitivity_LatCostScore_FailTime.csv};
\addplot [orange, thick, dashdotted , mark=diamond*] table [x index=7, y index=1, col sep=comma] {plots_tex/sensitivity_LatCostScore_FailTime.csv};
\addplot [green!70!black, thick, dashed , mark=x] table [x index=7, y index=2, col sep=comma] {plots_tex/sensitivity_LatCostScore_FailTime.csv};
\addplot [red, thick, dashdotted , mark=triangle*] table [x index=7, y index=3, col sep=comma] {plots_tex/sensitivity_LatCostScore_FailTime.csv};
\addplot [purple, thick, dashed , mark=+] table [x index=7, y index=4, col sep=comma] {plots_tex/sensitivity_LatCostScore_FailTime.csv};
\addplot [magenta, thick, solid, mark=o] table [x index=7, y index=5, col sep=comma] {plots_tex/sensitivity_LatCostScore_FailTime.csv};
\addplot [brown, thick, solid, mark=*] table [x index=7, y index=6, col sep=comma] {plots_tex/sensitivity_LatCostScore_FailTime.csv};

\end{axis}
\end{tikzpicture}
            \caption{Time to crash }            \label{fig:latcostfail}
        \end{subfigure}
    }

    \caption{Sensitivity analysis of the latency-cost score to varying anomaly rate and time to crash.}
    \label{fig:latcostsens}
\end{figure}
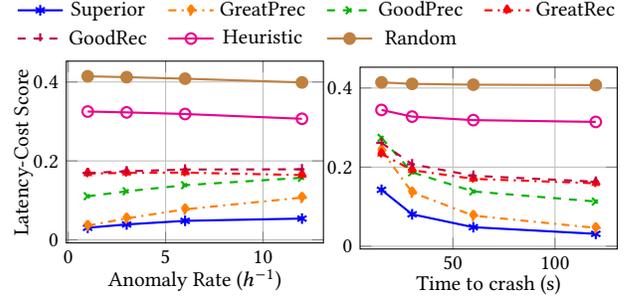

First, we test the detectors defined in Table~\ref{table:detecparams} with SRN variables from Table~\ref{table:setvars}, with different anomaly rates in $\{1h^{-1},$ $3h^{-1},$ $6h^{-1},$ $12h^{-1}\}$. Here, we consider a fixed inspection interval of $1s$. As shown in Figure~\ref{fig:latcostanom}, the overall comparison between detectors remains the same. Despite the anomaly rate impacting the average latency and resource consumption observed, the latency-cost score relation between detectors is not affected. Both latency and cost increase proportionally with anomaly rate, but this increase is similar on all tested detectors. The exceptions are detectors \textit{GreatPrec} and \textit{GoodPrec}, which present a lower recall than others, and so their score increases more with the anomaly rate.

In a second experiment, we vary the mean time it takes for a service to crash when in the anomalous state between values $\{15s,$ $30s,$ $60s,$ $120s\}$. These values correspond to the failure rates $\lambda_f$ of $\{240h^{-1},$ $120h^{-1},$  $60h^{-1},$ $30h^{-1}\}$, respectively. Again, the same detectors are tested with a fixed inspection interval of $1s$. We can observe in Figure~\ref{fig:latcostfail} that a shorter time to crash leads to an increase in the latency-cost score for \textit{Superior}, \textit{GreatPrec}, \textit{GoodPrec}, \textit{GreatRec} and \textit{GoodRec} detectors. This increase is mostly due to the greater chance of crashing and associated latency degradation. With a shorter time to crash, the advantage of precision over recall diminishes, as \textit{GreatPrec} and \textit{GoodPrec} yield results closer to those of \textit{GreatRec} and \textit{GoodRec}. This suggests that when there is limited time to detect an anomaly, both precision and recall become critical, as no other detector matches the trade-off achieved by \textit{Superior}.

It is worth pointing out that we do not evaluate the case of transient anomalies that do not cause crashes but still affect performance. Including such anomalies in the analysis would require a new model. This evaluation is left as future work.

\section{Related Work \label{sec:rwork}}
    
Models such as SRNs that represent the stochastic behavior of systems have been extensively used for Cloud performance and availability modeling. In \cite{ghosh2010end}, the authors leverage SRNs to analyze the impact of changes in workload (e.g., job arrival rate, job service rate), fault-load (e.g., physical machine failure rates) and system capacity on service unavailability and response delay. Our work presents a different analysis where we rather evaluate the impact of using an anomaly detector, to ensure good service performance. In \cite{machida2011candy}, the authors propose to construct analytic SRN models from existing system specification models for evaluating the availability of a Cloud web application. In \cite{longo2011scalable, GHOSH20131216}, the main contributions are on improving the scalability of performability analysis using SRNs of large-scale Cloud systems. These works focus on facilitating the modeling process of large Cloud environments for analysis. However, none of them model performance anomalies or an anomaly detection and correction process.

Stochastic models have also been used to analyze the outcomes of detections. The authors in \cite{andrade2019analysis} leverage a Deterministic and Stochastic Petri Net (DSPN) to quantitatively analyze the impacts of software aging in an anomaly detection process for a water treatment plant. Differently from our work, they analyze the degradation of the performance of anomaly detections, whereas we study the impact of anomaly detection on service performance and cost. In \cite{carnevali2024cost}, a Markov Regenerative Processes (MRGPs) approach is used to model different software rejuvenation policies for Software-Defined Networking (SDN) systems. The authors then analyze the impact of precision, recall, and inspection frequency of rejuvenation policies on system reliability and availability. Despite the similarities with our study, their work focuses on a different kind of system (SDN instead of Cloud services) and another type of anomaly (software aging instead of performance anomalies from antagonist jobs). 

Different works propose practical solutions for the detection and correction of anomalies in Cloud environments~\cite{chow2022deeprest, gan2021sage, ghosh2024fast, li2021multivariate}. In such works, the independent impact of recall, precision and inspection frequency of detectors on the monitored systems is hardly evaluated. This makes it difficult to tune the detector for specific needs, such as in the case of latency-critical services, where performance becomes crucial. Additionally, some of the most recent solutions use highly complex machine learning models to achieve high accuracy when detecting anomalies~\cite{gan2021sage, li2021multivariate}. However, high complexity can be an issue if it requires long processing times. As indicated by our evaluations, performing inspections in small intervals is important to achieve the best trade-off between performance and cost of the monitored system.

\section{Conclusion \label{sec:concl}}

This paper presents a study of the performance and cost implications of the properties of an anomaly detector in terms of precision, recall, and detection frequency on a Cloud service.
Using Stochastic Reward Nets, we simulate the usage of an anomaly detector to detect performance anomalies and trigger corrective scaling-out actions. By comparing different anomaly detector flavors, we evaluate the impact of their features on service latency and resource consumption. Our results show that both a high precision and recall are often not necessary for achieving a good trade-off between performance and cost. If the anomaly detector can be executed with small inspection intervals, then prioritizing precision allows for a close-to-optimal trade-off. Alternatively, if larger inspection intervals have to be used, prioritizing recall is more effective. 

Future extensions of this work may include expanding the model to multiple Cloud services, so that we can study the effects of an anomalous or crashed service on others. Another possible direction for further improvements would be to model other types of corrective action, such as scaling up or service migration.

\begin{acks}

This work was funded by the Region Auvergne-Rhône-Alpes through the Twin4FT project, the IPCEI-CIS E2CC project co-funded by BPI France under France 2030 and the European Commission for Next Generation Cloud Infrastructure and Services, and an IDEX Formation grant supporting collaboration between the University of Tsukuba and University Grenoble-Alpes.
\end{acks}

\bibliographystyle{ACM-Reference-Format}
\bibliography{bibliography}

\end{document}